\documentclass[iop]{emulateapj}
\usepackage{amsmath}

\shorttitle{Spin Axis Dynamics of a Moonless Earth}
\shortauthors{G. Li and K. Batygin }

\begin{document}
\newcommand{\icarus}{Icarus}

\title{On the Spin-axis Dynamics of a Moonless Earth}
\author{Gongjie Li \altaffilmark{1} and Konstantin Batygin \altaffilmark{1}}
\affil{1. Harvard-Smithsonian Center for Astrophysics}
\email{gli@cfa.harvard.edu}


\altaffiltext{1}{1Harvard-Smithsonian Center for Astrophysics, The Institute for Theory and
Computation, 60 Garden Street, Cambridge, MA 02138, USA}


\begin{abstract}
The variation of a planet's obliquity is influenced by the existence of satellites with a high mass ratio. For instance, the Earth's obliquity is stabilized by the Moon, and would undergo chaotic variations in the Moon's absence. In turn, such variations can lead to large-scale changes in the atmospheric circulation, rendering spin-axis dynamics a central issue for understanding climate. The relevant quantity for dynamically-forced climate change is the rate of chaotic diffusion. Accordingly, here we reexamine the spin-axis evolution of a Moonless Earth within the context of a simplified perturbative framework. We present analytical estimates of the characteristic Lyapunov coefficient as well as the chaotic diffusion rate and demonstrate that even in absence of the Moon, the stochastic change in the Earth's obliquity is sufficiently slow to not preclude long-term habitability. Our calculations are consistent with published numerical experiments and illustrate the putative system's underlying dynamical structure in a simple and intuitive manner.
\end{abstract}


\section{Introduction}
With the exception of Venus and Mercury, all planets in our solar system have satellites. However, satellites that comprise a high mass ratio are apparently not very common. In the solar system, the Earth-Moon system is the only planet (not counting Pluto-Charon) where $m_{M}/m_{\oplus}$ is not negligible. Moreover, no compelling evidence has been found for exomoons around the observed exoplanets \citep{Kipping13b, Kipping13a}.  

The existence of satellites with high mass ratios may play a significant role in stabilizing the planet's obliquity. For instance, the Earth's obliquity is currently stable. However, if the Moon were removed, the Earth's obliquity would undergo chaotic variations \citep{Laskar93b, Neron97,Lissauer12}. Mars's satellites comprise a negligible fraction of Mars' mass, and Martian obliquity is thought to have been chaotic throughout the solar system's lifetime \citep{Ward73, Touma93, Laskar93a}.   

The stability of the obliquity is very important for climate variations, as obliquity changes affect the latitudinal distribution of solar radiation. For the case of Mars (an ocean-free atmosphere-ice-regolith system), the obliquity changes apparently result in drastic variations of atmospheric pressure by runaway sublimation of $CO_2$ ice \citep{Toon80, Fanale82, Pollack82, Francois90, Nakamura03, Soto12}. For Earth-like planets (planets partially covered by oceans) the change of climate depends on the specific land-sea distribution and on the position within the habitable zone around the star. In other words, while it is debatable whether the variation in obliquity truly renders a planet inhabitable, it is clear that the climate can change drastically as the obliquity varies \citep{Williams97, Chandler00, Jenkins00, Spiegel09}.

Although spin-axis chaos for a Moon-less Earth is well established, the rate of chaotic diffusion appears to be inhomogeneous in the chaotic layer. To this end, \citet{Laskar93b} used frequency map analysis and noted that Earth obliquity may exhibit large variations (ranging from 0 degree to about 85 degree), if there were no Moon. However, recently \citet{Lissauer12} used direct integration and showed that the obliquity of a moonless Earth remains within a constrained range between $-2$ Gyr to $2$ Gyr, and concluded that the chaotic variations of the Earth's obliquity and the associated climatic variations are not catastrophic\footnote{This finding is in fact consistent with the frequency map analysis of \citet{Laskar93b}}. Stated more simply, it is not only important to understand if the obliquity undergoes chaotic variations but also to understand how rapidly such variations occur, to obtain a handle on the climatic changes that govern the habitability of a given planet. Our goal here is to describe a framework for such an analysis. We adopt a perturbative approach to the problem, and calculate the characteristic Lyapunov timescale and the diffusion coefficient of the obliquity. With the Lyapunov timescale and the diffusion coefficient, one can estimate the range of the obliquity the planet may reach in a given time, and inform the climate change of the planet.

Our paper is structured as follows. In section 2, we delineate the perturbative model and lay out the inherent assumptions. In section 3, we calculate the  diffusive properties of the system and compare our analytical estimates to numerical simulations. We conclude and discuss the implications of our results in section 4.

\section{A Simplified Perturbative Model}
As the primary goal of this work is to obtain analytical estimates of the relevant timescales for chaotic diffusion, we begin by considering a simplified description of the system.

Without the Moon, the Earth's obliquity is found to be chaotic in the range $0-85^o$ , where there are two large chaotic regions: $0^o-45^o$ \& $65^o-85^o$. There also exists a moderately chaotic bridge that connects the two regions: $45^o-65^o$ \citep{Laskar93b, Morbidelli02}. The dynamical analysis is simpler in the large chaotic regions. Thus, we treat them first.

\subsection{Large chaotic regions: $0^o-45^o$ \& $65^o-85^o$}
The Hamiltonian describing the evolution of planetary obliquity is well documented in the literature (e.g. \citet{Colombo66, Laskar93a, Touma93, Neron97}):
\begin{eqnarray}
\label{eqn:HF}
H(\chi, \psi, t) &=& \frac{1}{2}\alpha\chi^2 +\sqrt{1-\chi^2} \\ \nonumber
						  &\times& (A(t)\sin{\psi}+B(t)\cos{\psi})),
\end{eqnarray}
where $\psi$ is the longitude of the spin-axis, $\chi = \cos \varepsilon$, $\varepsilon$ is the obliquity, and $\alpha$ is an approximately constant parameter. Specifically,
\begin{eqnarray}
\alpha &=& \frac{3G}{2\omega}\Big[\frac{m_{\odot}}{(a_{\odot}\sqrt{1-e_{\odot}^2})^3} \\ \nonumber
			&+&\frac{m_{M}}{(a_{M}\sqrt{1-e_{M}^2})^3}(1-\frac{3}{2}\sin^2i_M)\Big]E_d,
\end{eqnarray}
where $m_{\odot}$ is the mass of the Sun, $a_{\odot}$ and $e_{\odot}$ are the semi major axis and the eccentricity of the Earth's orbit, $m_{M}$ is the mass of the moon, $a_{M}$, $e_{M}$ and $i_{M}$ are the semi major axis, eccentricity and inclination of the Moon's orbit around the Earth, $E_d$ is the dynamical ellipticity of Earth, and $\omega$ is the spin of the Earth. $\alpha$ characterizes the intrinsic precession of the Earth's spin axis, and is obtained by averaging the torques from the Sun and Moon over their respective orbits. For a moonless Earth, $\alpha = 0.0001 \rm{yr} ^{-1}$ \citep{Neron97}. In addition,

\begin{eqnarray}
A(t) = 2(\dot{q}+p(q\dot{p}-p\dot{q}))/\sqrt{1-p^2-q^2}, \\
B(t) = 2(\dot{p}-q(q\dot{p}-p\dot{q}))/\sqrt{1-p^2-q^2},
\end{eqnarray}
where $p=\sin{i/2}~\sin{\Omega}$ and $q=\sin{i/2}~\cos{\Omega}$, $i$ is the inclination of the Earth with respect to the fixed ecliptic and $\Omega$ is the longitude of the node. 

The inclination and the longitude of node of the Earth change as the other planets in the solar system perturb the Earth's orbit. The evolution of $i$ and $\Omega$ can be obtained by direct numerical integration or in the low-$e$,$i$ regime via perturbative methods such as the Lagrange-Laplace secular theory. Specifically, within the context of the latter, a periodic solution represented by a superposition of linear modes can be obtained. 

\begin{eqnarray}
\label{eqn5}
i\cos{\Omega} = \sum i_k\cos{(s_k t + \gamma_k)}, \\
\label{eqn6}
i\sin{\Omega} = \sum i_k\sin{(s_k t + \gamma_k)}.
\end{eqnarray}

The amplitudes and the frequencies of the modes have been computed in classic works \citep{LeVerrier55, Brouwer50}. We use the latest update of these values from \citet{Laskar90}.

In adopting equations (\ref{eqn5}) and (\ref{eqn6}) as a description of the EarthÕs inclination dynamics, we force the disturbing function in Hamiltonian (1) to be strictly periodic. In fact, it is well known that the orbital evolution of the terrestrial planets is chaotic with a characteristic Lyapunov time of $\sim 5$Myr \citep{Laskar89, Sussman92}. Consequently, our model does not account for the stochastic forcing of the obliquity by the diffusion of the EarthÕs inclination vector \citep[see][]{Laskar93b}. Such a simplification is only appropriate for systems where the intrinsic Chirikov diffusion is faster than that associated with the disturbing function. As will be shown below, the assumption holds for the system at hand. 

\begin{figure}[h]
\includegraphics[width=3in, height=1.85in]{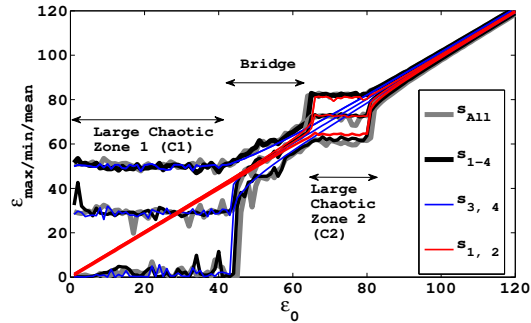}
\caption{\label{f:4freq} The minimum/mean/maximum of the obliquity reached in 18Myr as a function of the initial obliquity. The grey lines represent the results including all the frequencies, the black lines represent the results including $s_1$, $s_2$, $s_3$ and $s_4$, the red lines represent the results including $s_1$ and $s_2$, and the blue lines represent the results including $s_3$ and $s_4$. The four frequencies reproduces the results including all the frequencies. Between $\varepsilon_0 \sim 65^o - 85^o$ and $\sim 0^o - 45^o$, the chaotic behavior of obliquity is caused by $s_1$ and $s_2$, and $s_3$ and $s_4$ separately. Between $45^o - 65^o$, the evolution of the obliquity is also not regular, and is caused by a nonlinear coupling among the resonant doublets ($s_{1, 2}$ and $s_{3, 4}$).}
\vspace{0.1cm}
\end{figure}

As already mentioned above, in absence of the Moon, rapid chaos spans two well-separated regions, which are joined by a weakly chaotic bridge \citep{Laskar93b}. In each of the highly chaotic regions, irregularity arises from overlap of a distinct pair of secular resonances (see \citet{Chirikov79}). As shown in Figure (\ref{f:4freq}), the overlap of $s_1$ and $s_2$ causes the chaotic region in $\varepsilon \sim 65^o - 85^o$ (``C2") and the overlap of $s_3$ and $s_4$ causes the chaotic region in $\varepsilon \sim 0^o - 45^o$ (``C1"). Including only the terms associated with these four frequencies in Hamiltonian (\ref{eqn:HF}), the chaotic region of $\varepsilon \sim 0^o - 85^o$ can be well reproduced \citep{Morbidelli02}. Accordingly in the following analysis, we retain only the four essential modes to analyze the two chaotic regions and the ``bridge" that connects them sequentially. 

Substituting the expansion for $i\cos{\Omega}$ and $i\sin{\Omega}$ and keeping only the four frequencies ($s_{1-4}$), we can rewrite the Hamiltonian as 
\begin{eqnarray}
\label{eqn:HO1}
H_{C1, 2} (\chi, \psi, t) &=& \frac{\alpha}{2} \chi^2 + \epsilon \sqrt{1-\chi^2} \\ \nonumber
											&\times&	\sum_{k = 1}^4 a_k \cos{( s_k t + \delta_k + \psi )},
\end{eqnarray}
where $\epsilon = 10^{-7}$, $\alpha = 0.0001 \rm{yr}^{-1}$. The other parameters are included in table (\ref{t:O1}).

\begin{table}
\caption{Parameters for the simplified Hamiltonian (\ref{eqn:HO1}).}
\label{t:O1}
  \centering
	\begin{tabular}{ |l|l|l|l| }
    \hline
       & a ($\rm{yr} ^{-1}$) & s ($\times10^{-5} \rm{yr} ^{-1}$)& $\delta$ \\ \hline
    k = 1 & 2.47638 &  -2.72353 & -2.56678\\ 
    k = 2 & 2.93982 & -3.43236 & -1.70626\\
    k = 3 & 15.5794 & -9.1393 & 1.1179\\ 
    k = 4 & 5.46755 & -8.6046 & 2.4804\\
    \hline
	\end{tabular}
\end{table}

\begin{figure}[h]
\includegraphics[width=3in, height=1.9in]{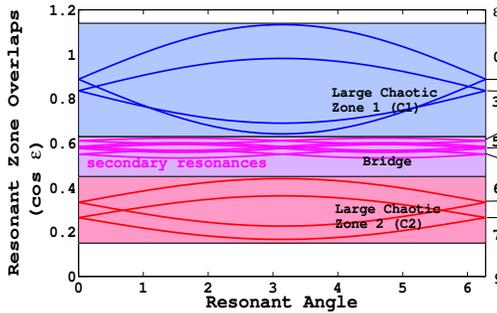}
\caption{\label{f:resolap} The overlap of primary and secondary resonances. The red lines represent the resonances of $s_1$ and $s_2$, while the blue lines represent the resonances of $s_3$ and $s_4$. The purple lines represent the second order resonances in the bridge. }
\vspace{0.1cm}
\end{figure}

Within ``C1" and ``C2", the chaotic variations are not sufficiently large to induce overwhelming variations in $\sqrt{1-\chi^2}$. To leading order, it can be assumed to be constant, and we evaluate it at the center of the chaotic regions (specifically $\chi_{0, 1}= 20^\circ$ for ``C1" and $\chi_{0, 2} = 70^\circ$ for ``C2"). In doing so, we deform the topology of the domain inherent to Hamiltonian (7) from a sphere to a cylinder. While not appropriate in general, such an operation is justified for the system at hand because both ``C1" and ``C2" individually occupy a limited obliquity range (see Appendix for additional discussion). Then, the Hamiltonian obtains a simple pendulum like structure, characterized by four forced resonances. Keeping one resonance at a time, we can plot the separatrixes associated with each harmonic (Figure (\ref{f:resolap})). As noted before, the two large chaotic zones can be understood to be the interaction of the resonant doublets  $s_1$ \& $s_2$, and $s_3$ \& $s_4$ separately. The region in the bridge is dominated by the secondary resonances which will be described in the next section. We also note that by setting $\sqrt{1-\chi^2}$ to a constant, the ``C1'' region extends to $\chi = \cos{\varepsilon} >1$. Because here we only focus on the qualitative dynamical behavior, the extension to the unphysical regions can be neglected. Furthermore, we notice that there is a gap between the second order resonances and the ``C2" region. This gap is likely also an artifact that arises from setting $\sqrt{1-\chi^2} = const.$, as this assumption leads to a deformation of the resonant structure. Since the dynamical behavior in the bridge is well characterized by a second-order truncation of the averaged Hamiltonian, we do not extend our analysis to the higher orders. 

Keeping $s_{3, 4}$ or $s_{1, 2}$ only, we can adequately reproduce the large chaotic regions ``C1" and ``C2". Thus, we obtain simplified Hamiltonians for ``C1" and ``C2" separately. 

\begin{eqnarray}
\label{eqn:C1}
H_{C1} (\chi, \psi, t) &=& \frac{\alpha}{2} \chi^2 + \epsilon \sqrt{1-\chi_{0, 1}^2} \\ \nonumber
									&\times&( a_3 \cos{( s_3 t + \delta_3 + \psi )} \\ \nonumber
										&+&a_4 \cos{( s_4 t + \delta_4 + \psi )}),			
\end{eqnarray}

\begin{eqnarray}		
\label{eqn:C2}						
H_{C2} (\chi, \psi, t) &=& \frac{\alpha}{2} \chi^2 + \epsilon \sqrt{1-\chi_{0, 2}^2} \\ \nonumber
											&\times&( a_1 \cos{( s_1 t + \delta_1 + \psi )} \\ \nonumber
											&+&a_2 \cos{( s_2 t + \delta_2 + \psi )}),
\end{eqnarray}
where the parameters can be found in table (\ref{t:O1}).

\subsection{Bridge region: $45^o-65^o$}
In case of the Earth, if the obliquity were to be confined to either large chaotic domain, the climatic variability could in principle be relatively small. However, the analysis of \citet{Laskar93b} shows that transport between the two regions is possible. To understand the migration between the two chaotic zones, one needs to understand the dynamics in the bridge zone between $45^o - 65^o$. As the bridge zone is a region between the two doublet resonant domains, it is likely that the diffusion is driven by secondary, rather than primary resonances. In this section, we present the simplified Hamiltonian governing the dynamical behavior in the bridge. 

In order to obtain an analytical description of the resonant harmonics in the bridge, we must generate them by averaging over the primary harmonics. In particular, here we do so by utilizing the Poincare-Von Ziepel perturbation method \citep{Goldstein50, Lichtenberg83}. Consider a canonical transformation that arises from a type-2 generation function $G(\Phi, \psi, t)$, $\Phi = \chi - \epsilon \frac{\partial G}{\partial \psi}$, $\phi = \psi +\epsilon \frac{\partial G}{\partial \Phi}$. Upon direct substitution, we obtain:

\begin{eqnarray}
\label{eqn:HO2}
H_B (\Phi, \phi, t) &=& \frac{\alpha}{2}\Phi^2 + \epsilon^2 \Big[ \frac{\alpha}{2}\Big(\frac{\partial G}{\partial \psi}\Big)^2\\ \nonumber
&+&\sum_{i=1}^4 \sin(s_i t + \phi) \frac{\partial G}{\partial \Phi} \Big] + O(\epsilon^3),
\end{eqnarray}
where
\begin{eqnarray}
\label{eqn:G}
G (\Phi, \psi, t) &=& \frac{\sqrt{1-\Phi^2}}{\prod_{k=1}^4 (\alpha\Phi+s_k)} \Big(\\
							&-&	\sum_{i=1}^4 \alpha^3\Phi^3\sin(\psi+s_i t+\delta_i)a_i \nonumber \\
							&-& \sum_{j\neq i} \alpha^2\Phi^2\sin(\psi+s_i t+\delta_i)a_i s_j \nonumber \\
							&-& \sum_{j, l\neq i} \alpha\Phi\sin(\psi+s_i t+\delta_i)a_i s_js_l \nonumber \\
							&-& \sum_{j, l, m\neq i} \sin(\psi+s_i t+\delta_i)a_i s_js_l s_m\Big). \nonumber
\end{eqnarray}

As before, we set $\Phi$ to a constant (at $\Phi=\cos(50^o)$) in the second term and rewrite the Hamiltonian as:
\begin{eqnarray}
H_{B}(\Phi, \phi, t) &=& \frac{\alpha}{2} \Phi^2 \\ \nonumber
									&+& \epsilon^2 \Big(\sum_k b_k \cos{( s_{2,k} t + \delta_{2, k} + 2\phi )}\Big),
\end{eqnarray}
where $s_{2, k}$ is the sum of any two of the first order resonance frequencies $s_{1-4}$. Note that because the bridge region is even more tightly confined in obliquity than either ``C1" or ``C2", it is sensible to ignore the variations in $\Phi$ in the second term.

Considering each resonant term in isolation, the Hamiltonian resembles that of a simple pendulum. Plotting the separatrix of the Hamiltonian for each term, we find that there are four second order resonances in the bridge region (as shown in Figure (\ref{f:resolap})). Two of the resonances reside in extreme proximity to each-other and only give rise to modulational diffusion that is much slower than that arising from marginally overlapped harmonics \citep{Lichtenberg83}. Consequently, we can approximate the Hamiltonian in the bridge by three overlapping resonances. Thus, the simplified Hamiltonian for the bridge is:
\begin{eqnarray}
\label{eqn:SHO2}
H_B(\Phi, \phi, t) &=& \frac{\alpha}{2} \Phi^2 \\ \nonumber
								&+& \epsilon^2 \sum_{k=1}^3 b_k \cos{( s_{2,k} t + \delta_{2, k} + 2\phi )},
\end{eqnarray}
where the parameters can be found the table (\ref{t:O2}).

\begin{table}[h]
	\caption{Parameters for the simplified Hamiltonian (\ref{eqn:SHO2}).}
	\label{t:O2}
  \centering
	\begin{tabular}{ |l|l|l|l| }
    \hline
       & b  ($\rm{yr} ^{-1}$) & s  ($\rm{yr} ^{-1}$) & $\delta$ \\ \hline
    k=1 & 789482. & $s_1+s_3 = -0.000118628$ & -1.69271\\ 
    k=2 & 755727. & $s_1+s_4= -0.000113281$ & -3.05521\\
    k=3 & 364558. & $s_2+s_3 = -0.000125717$ & -2.55324\\ 
   \hline
	\end{tabular}
\end{table}


\section{Results}
\subsection{Analytical Estimates}
With simplified expressions for the Hamiltonians in each charscteristic region delineated, we can estimate quantities relevant to the extent of the motion's irregularity (specifically, the Lyapunov exponent and the action-diffusion coefficient) following \citet{Chirikov79}, with the method discussed in the Appendix. 

Briefly, for a simple asymmetrically modulated pendulum:
\begin{eqnarray}
H_D(p, q, t) = \frac{\beta}{2}p^2 + c (\cos q + \cos(q+\omega_B t)),
\end{eqnarray}
where there are two resonant regions separated by $\omega_B/\beta$ in action. The libration frequency associated with the stable equilibria of either resonance is $\omega_L = \sqrt{c\beta}$, which is identical (in magnitude) to the unstable eigenvalue of the separatrix. Moreover, the half width of the resonance, $\Delta$, can be calculated as $\Delta = 2\sqrt{c/\beta}$. 

When the resonances are closely overlapped (e.g. in region ``C1", ``C2"), the Lyapunov exponent ($\lambda$) is roughly the breathing frequency: $\nu_B = \omega_B/(2\pi)$. Meanwhile, in the marginally overlapped case (``bridge"), it amounts to roughly $2\nu_L = 2\omega_L/(2\pi)$. In other words,
\begin{eqnarray}
\lambda \sim \frac{1}{K} \frac{\omega_L}{2\pi} \sim \left\{ \begin{array}{rl}
 \nu_B &\mbox{($\omega_B / \beta < \Delta$)} \\
 2\nu_L &\mbox{($\omega_B / \beta \sim 2 \Delta)$},
       \end{array} \right.
\end{eqnarray}
where $K = \frac{\Delta}{\omega_B/\beta} = 2\frac{\omega_L}{\omega_B}$ is a stochasticity parameter, which characterizes the extent of resonance overlap. Note that when $\omega_B < \Delta$, $\frac{1}{K} \frac{\omega_L}{2\pi} = \nu_B/2$. We adopt $\lambda \sim \nu_B$ based on the results from the Appendix in the following calculation. The empirical factor of 2 does not affect our results on the qualitative behavior of the system.

Accordingly, the quasi-linear diffusion coefficient ($D$) can be estimated as $\Delta^2\nu_B$ when the resonances are closely overlapped, and as $\Delta^2\nu_L$ when the resonances are farther apart \citep{Murray85}, although better estimates can be obtained in adiabatic systems \citep{Cary86, Bruhwiler89, Henrard93}:
\begin{eqnarray}
D \sim \Delta^2\lambda \sim \left\{ \begin{array}{rl}
  \Delta^2\nu_B &\mbox{($\omega_B/\beta < \Delta$)} \\
  \Delta^2\nu_L &\mbox{($\omega_B/\beta \sim 2\Delta$)}.
       \end{array} \right.
\end{eqnarray}

Taking the simplified Hamiltonian (\ref{eqn:C1}) and following \citet{Murray97}, we approximate the two resonances as having the same widths (which quantitatively amount to the average width). Upon making this approximation, we get:
\begin{eqnarray}
\tilde{H}_{C1} (\chi, \psi, t)&=& \frac{\alpha}{2} \chi^2 +\epsilon\tilde{a}_2(\cos (s_3t+\delta_3+\psi) \nonumber \\
										&+& \cos{(s_4t +\delta_4+\psi)}).
\end{eqnarray}
As noted earlier, $\epsilon = 10^{-7}$, $\alpha = 0.0001 \rm{yr} ^{-1}$, $s_4-s_3 = 5\times10^{-6} \rm{yr} ^{-1}$, $\tilde{a}_1 = 3.6  \rm{yr} ^{-1}$. 

Because the two resonances are closely overlapped (as shown in Figure (\ref{f:resolap})), the Lyapunov exponent can be estimated as the breathing frequency: $\nu_B=(s_4-s_3)/2/\pi \sim 10^{-6} \rm{yr} ^{-1}$. Accordingly, the diffusion coefficient ($D$) is $\Delta^2 \nu_B \sim 10^{-8} \rm{yr} ^{-1}$, where $\Delta = 2\sqrt{\epsilon\tilde{a}/\alpha}$ is the half width of the resonant region. With the diffusion coefficient, we can estimate the time needed to cross the two chaotic zones and the bridge: $t \sim \delta \chi^2 /D$. Specifically, taking $\delta \chi = \cos{0^o}-\cos{45^o}$, $t_{C1} \sim 7.5$ Myr.

Next, we consider zone ``C2" (equation (\ref{eqn:C2})). After approximating the two resonances as having the same width, we rewrite the Hamiltonian as
\begin{eqnarray}
\tilde{H}_{C2} (\chi, \psi, t)&=& \frac{\alpha}{2} \chi^2 + \epsilon \tilde{a}_2(\cos(s_1t + \delta_1 +\psi) \nonumber \\
										&+&\cos(s_2t +\delta_2+\psi)),
\end{eqnarray}
where $\alpha = 0.0001 \rm{yr} ^{-1}$, $\epsilon = 10^{-7}$, $\tilde{a}_2 = 2.5 \rm{yr} ^{-1}$, $s_1-s_2 = 7\times10^{-6} \rm{yr} ^{-1}$. 

Similarly to zone ``C1", the Lyapunov exponent can be estimated as $\nu_B \sim 10^{-6} \rm{yr} ^{-1}$, because the two resonance are closely overlapped. The diffusion coefficient thus evaluates to $D = \Delta^2 \nu_B \sim 10^{-8} \rm{yr} ^{-1}$. Finally, the time to cross ``C2" can be estimated as $t_{C2} \sim \delta \chi^2 / D \sim 10$ Myr.

Finally, for the bridge zone, we can approximate the simplified Hamiltonian in equation (\ref{eqn:SHO2}) as a resonance triplet with the same width: 
\begin{eqnarray}
\tilde{H}_{B} (\Phi, \phi, t)&=& \frac{\alpha}{2} \Phi^2 \\ \nonumber
						&+&\epsilon^2\tilde{a}_3(\cos (s_{2, 1} t + \delta_{2, 1} + 2\phi ) \\ \nonumber
						&+&\cos{ (s_{2, 2} t + \delta_{2, 2} + 2\phi )} \\ \nonumber 
						&+&\cos{ (s_{2, 3} t + \delta_{2, 3} + 2\phi )}),
\end{eqnarray}
where $\alpha = 0.0001  \rm{yr} ^{-1}$, $\epsilon = 10^{-7}$, $\tilde{a}_3 = 664633  \rm{yr} ^{-1}$, $\delta_s = 6.21769\times10^{-6}$. 

Because the resonances are not closely overlapped as shown in Figure (\ref{f:resolap}), the Lyapunov exponent can be estimated as $2\omega_L/(2\pi)$, where the libration frequency is $\omega_L = \sqrt{2\alpha(\epsilon^2\tilde{a})}$ (the angle is $2\phi$ instead of $\phi$). Thus, the Lyapunov exponent is roughly $\sim 3.7\times10^{-7} \rm{yr} ^{-1}$. Then the diffusion coefficient can be estimated as $\Delta ^2 \nu_L \sim 5\times10^{-11} \rm{yr} ^{-1}$, and $t_{bridge} \sim 2$ Gyr. 

The stark differences in the estimates of the crossing times obtained above place the results of \citet{Lissauer12} into a broader context. That is, our calculations explicate the fact that the long-term confinement of the obliquity to either the ``C1" or the ``C2" regions observed in direct numerical simulations arises from the distinction in the underlying resonances that drive chaotic evolution. Because the diffusion in the bridge is facilitated by secondary resonances, it is considerably slower, allowing the stochastic variation in obliquity to remain limited.  

\subsection{Numerical Results}
To validate the analytical results, we numerically estimate the Lyapunov exponent. We follow the method discussed in Ch. 5 of \citet{Morbidelli02}. Specifically, we linearize the Hamiltonian and evolve the difference ($\delta_{traj} (t)$) of two initially nearby trajectories in phase space. The initial separation is set to $10^{-6}$. The Lyapunov exponent is calculated as:
\begin{eqnarray}
\lambda = \displaystyle \lim_{t \to \infty} \frac{1}{t}\ln\frac{\delta_{traj} (t)}{\delta_{traj} (0)}.
\end{eqnarray}

\begin{figure}[h]
\includegraphics[width=3in, height=1.9in]{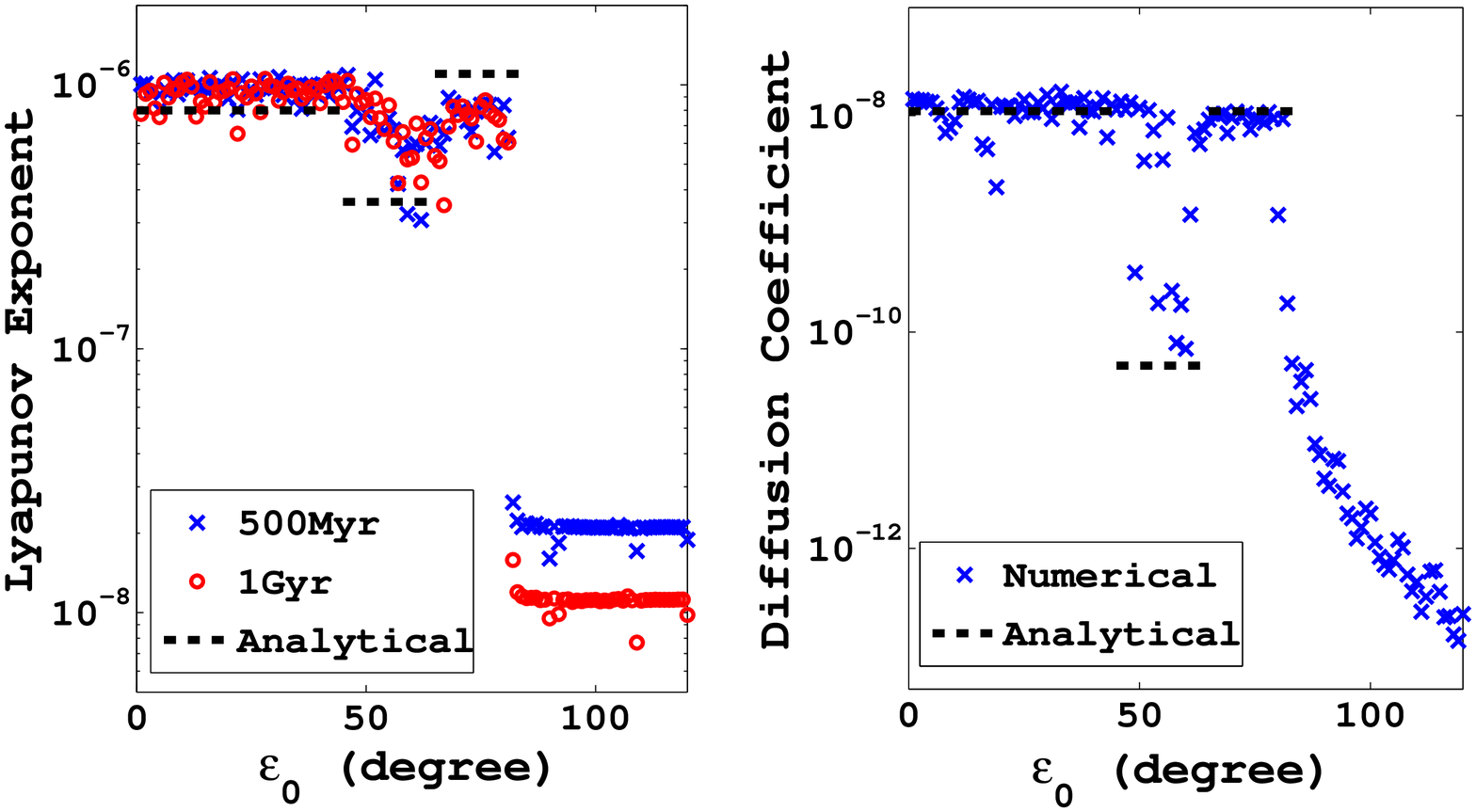}
\caption{\label{f:Nresult} The numerical result of the Lyapunov exponent and the diffusion coefficient with different initial obliquity. Left panel: the Lyapunov exponent. The red circles represent the Lyapunov exponent calculated for $t = 500$ Myr, and the blue crosses represent that calculated for $t = 1$ Gyr. The Lyapunov exponent converges in the chaotic region for the different running times and in the regular region the Lyapunov exponent approaches zero as the running time increases. Right panel: the diffusion coefficient estimated by taking averages over bins of $0.5$ Myr before taking the difference in $\chi$. The diffusion coefficient in the bridge is much smaller than that in the chaotic zones. The dashed lines in the two panels are the results using the analytical method.}
\vspace{0.1cm}
\end{figure}

We start our runs with different initial obliquity to probe the different chaotic/regular regions. We check the convergence of our results using two different running times ($t = 500$ Myr and $t = 1$ Gyr). In the regular regions, the Lyapunov exponent approaches zero, and is limited only by the integration time. As shown in Figure (\ref{f:Nresult}), the Lyapunov exponents in the two large chaotic zones are $\lambda_{C1}\sim\lambda_{C2}\sim 10^{-6} \rm{yr} ^{-1}$ and the Lyapunov exponents in the bridge zone is $\lambda_{bridge}\sim 5\times10^{-7} \rm{yr} ^{-1}$.

Then, we follow the numerical method discussed in \citet{Chirikov79} to calculate the diffusion coefficient. Specifically, to eliminate the oscillations caused by the libration of the resonances, we average $\chi$ in bins with the same bin size $\delta t$. Then, we take the difference ($\delta \chi$) between neighboring bins. The diffusion coefficient is estimated by averaging $\delta \chi^2 / \delta t$. The bin size $\delta t$ needs to be bigger than the libration period of the resonances but smaller than the saturation timescale in the chaotic zone and the bridge. Here, we set $\delta t = 0.5$ Myr, and run the simulation for 500 Myr. The results are plotted in the right panel of Figure (\ref{f:Nresult}). Unsurprisingly, the diffusion coefficient is much smaller in the bridge than that in the chaotic zones. 

\begin{figure}[h]
\includegraphics[width=3in, height=1.9in]{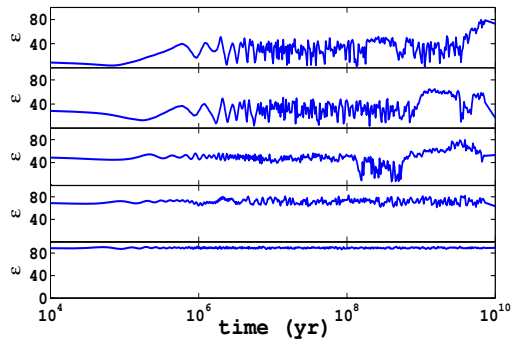}
\caption{\label{f:fullrun} The evolution of the obliquity as a function of time by integrating the full secular Hamiltonian numerically (equation (\ref{eqn:HF})). The different panels represent different initial obliquities: from top to bottom: $\varepsilon_0 = 10^o$, $\varepsilon_0 = 30^o$, $\varepsilon_0 = 50^o$, $\varepsilon_0 = 70^o$. $\varepsilon_0 = 90^o$. $\psi_0 = 0$ for all the panels. }
\vspace{0.1cm}
\end{figure}

We compare the analytical results with the numerical estimation. In Figure (\ref{f:Nresult}), the analytical results are represented by black dashed lines. Roughly, the analytical results are consistent with the numerical results. To further elucidate the qualitative agreement, we integrated the full Hamiltonian (equation (\ref{eqn:HF})) and the resulting evolutionary sequences are shown in Figure (\ref{f:fullrun}).

Note that the time to cross ``C1" and ``C2" are about $\sim$few Myr, and the time to cross the bridge is much longer: $\gtrsim$ Gyr. This is fully consistent with our analytical arguments. Furthermore, as already mentioned above our results are consistent with \citet{Lissauer12}, who noticed that the Earth's obliquity is constrained in ``C1" within $-2$ Gyr to $2$ Gyr. Although the diffusion time we calculated for the bridge is $\sim2$ Gyr, the diffusion time only roughly characterizes the timescale it takes to cross the bridge, and the exact crossing time depends on the specific initial condition. Thus, as $2$ Gyr is on the similar timescale of the integration time used in \citet{Lissauer12}, it is probable that the obliquity would reach ``C2" if the integration time in their simulations were to be increased.

\section{Conclusion}
Without the Moon, Earth's obliquity is chaotic, however, the rate at which the system explores the irregular phase space is not evident a-priori \citep{Laskar93b, Lissauer12}. In other words, the characteristic range over which the obliquity varies in a given time-frame depends sensitively on the exact architecture of the underlying resonances that drive chaotic motion. Here, we utilized canonical perturbation theory to estimate the Lyapunov exponent and the diffusion coefficient which characterize the chaotic rate of the change of the obliquity. Our calculations were performed within the context of a perturbative approach which yields a simple model, which in turn illuminates the underlying structure of the dynamics in a direct and intuitive way.

In order to obtain a qualitatively tractable description of the system, we simplified the Hamiltonian to a restricted sum of single pendulums, and followed \citet{Chirikov79} to estimate the characteristic timescales. Subsequently, we validated the analytical results by calculating the Lyapunov exponent and diffusion coefficient numerically and by integrating the full Hamiltonian in the secular approximation. We found broad agreement between the analytical and numerical results. Particularly, there are three distinct regions where the obliquity exhibits chaotic variations. Rapid chaos is observed between $0-45^o$ and $65-85^o$, while a mildly chaotic bridge connects the two regions. Our estimates suggest that the time to cross the ``bridge" is $\sim 2$ Gyr, much longer than the time to cross the two large chaotic zones. This is consistent with the findings of \citet{Lissauer12}. 

With the envelope of the exoplanetary detection edging ever closer to the discovery of numerous Earth-like planets\footnote{To date, the recently completed Kepler mission has detected four super-Earths (namely Keper-22b,62e,62f,69c) in the habitable zone \citep{Borucki12, Borucki13, Barclay13}.}, the spin-axis dynamics of a Moonless Earth presents itself as an important paradigm for the assessment of the potential climate variations on such objects. Indeed, it is tempting to apply a framework such as that outlined in this work to an array of multi-transiting planetary systems, for which the masses and orbital parameters are well established. Unfortunately, results stemming from such an exercise would be under-informed by a lack of observational constraints on the physical properties of the individual planets such as spin rates and dynamical ellipticities. Consequently, endeavors of this sort must await substantial breakthroughs in observational characterization. Nevertheless, the implications of the present study for the emerging extrasolar planetary aggregate are clear: an absence of a high-mass ratio Moon should not be viewed as suggestive of extreme climate variations. That is, even for a Moonless Earth-like planet, residing in a stochastic spin-axis state, the characteristic chaotic diffusion rate may sufficiently slow to not limit long-term habitability.

\acknowledgments

\bibliographystyle{hapj}
\bibliography{msref}

\appendix
\subsection{Dynamics of the Unsimplified Hamiltonian}
In our simplified perturbative model, we set $\sqrt{1-\chi^2}$ to be a constant in the Hamiltonian (equation (\ref{eqn:HO1})) in order to treat this system as a modulated pendulum. Here, we justify this approach by showing that the dynamics with the original Hamiltonian can be well approximated by the simplified version with $\sqrt{1-\chi^2}$ set to be a constant. 

\begin{figure}[h]
\center
\includegraphics[width=5in, height=2.3in]{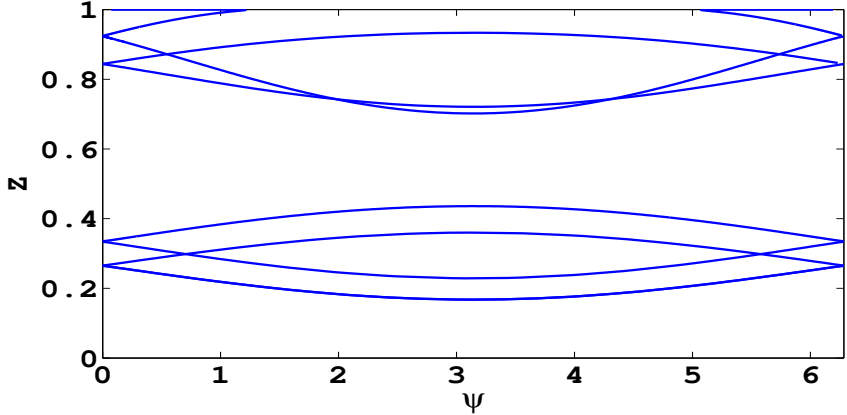}
\caption{\label{f:Col} 
The separatrix of the un-simplified Hamiltonian with each frequency $s_k$. It is analogous to that in Figure \ref{f:resolap}, justifying our approaching with $\sqrt{1-\chi^2}$ set to be a constant.}
\vspace{0.1cm}
\end{figure}

Considering each forcing term with frequency $s_k$ at a time, we can plot the critical curve of the trajectories. We show the four critical curves with the different frequency $s_k$ in Figure \ref{f:Col}. Since most of the forcing terms have librating region far from $\chi = 1$, the separatrixes are not greatly distorted and are essentially analogous to that with $\sqrt{1-\chi^2}$ constant in Figure \ref{f:resolap}. Because the interaction of the resonant (librating) regions give rise to the dynamical structure of this system, the corresponding overlaps of the separatrixes demonstrate that the dynamics of the original Hamiltonian can be captured by the simplified Hamiltonian. 

\subsection{Double Resonances and Triple Resonances}
As explained in the main text, the chaotic zones and the bridge can be approximated as two or three overlapping resonances with equal widths. Here, we demonstrate an analytical way to calculate the Lyapunov exponent and the diffusion coefficient for the double or triple resonances with the same resonant widths. This analytical method can be applied for resonant doublets or triplets with equal widths in general. 

\begin{figure}[h]
\includegraphics[width=6in, height=2.3in]{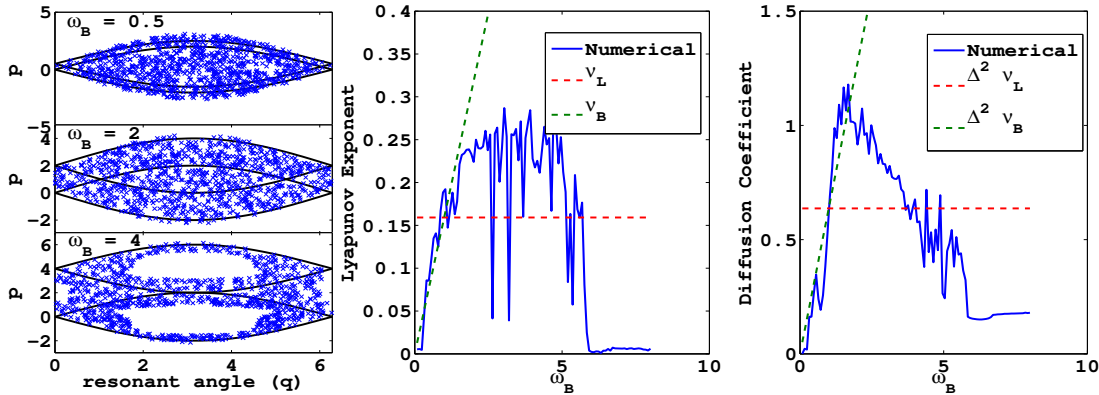}
\caption{\label{f:double} The analytical models for the double resonances. Left panel: the surface of section of the double resonances with different overlaps starting with $q=1$, $p=1.5$. Middle panel: the numerical and the analytical estimates of the Lyapunov exponent with different overlaps. Right panel: the numerical and analytical estimates of the diffusion coefficient with different overlaps.}
\vspace{0.1cm}
\end{figure}

For the double resonances, the Hamiltonian can be written as:
\begin{eqnarray}
H_D(p, q, t) = \frac{\beta}{2}p^2 + c (\cos q + \cos(q - \omega_B t))
\end{eqnarray}
where $\omega_B$ is the frequency difference between the two resonances. The half width of each resonant region is $\Delta = 2\sqrt{c/\beta}$, and the libration frequency of each resonant region is $\omega_L = \sqrt{c\beta}$. To illustrate the behavior of this Hamiltonian, in Figure (\ref{f:double}),  we plot the surface of section starting from point $p=1.5$, $q=1$ with the total run time $t=1000$, where we measure time in units of $1/\sqrt{c \beta}$ and action in units of $\sqrt{c/\beta}$. As $\omega_B$ decreases, the two resonances are more overlapped. 
 
Next, we estimate the Lyapunov exponent of the double resonances numerically. Following the method discussed in \citep{Morbidelli02}, we linearized the hamiltonian $H_D$ to evolve the difference of two trajectories. We start the integration at $p=1.5$, $q=1$ arbitrarily, and calculate the Lyapunov exponent as $\frac{1}{t}\ln\frac{\delta (t)}{\delta (0)}$, where we set $t=1000$ for our integration. We plot the numerical result in the middle panel of Figure (\ref{f:double}) with the blue line. To compare with the characteristic frequencies in this system, we over plotted $\nu_B = \omega_B/2\pi$ and $\nu_L = \omega_L/2\pi$. 

We notice that when the resonances are closely overlapped $\omega_B<2$, the Lyapunov exponent can be approximated as $\nu_B$. When the resonances are less overlapped but still attached $2<\omega_B<4$, the Lyapunov exponent is approximately constant ($\sim 2 \omega_L$). When the resonances are more separated, the Lyapunov exponent falls as the system becomes more regular. 

Then, we calculate the diffusion coefficient numerically. To average over the oscillations due to the libration behavior, we take the difference in $\delta p$ at $t = n /\nu_B$, $n \in Z$, and estimate the diffusion coefficient as $\langle \delta p^2 \nu_B \rangle$. The result is plotted in the right panel in Figure (\ref{f:double}) with the blue line. 

Comparing with the characteristic timescale of the system, we find that the when the two resonant regions are closely overlapped ($\omega_B<2$), the diffusion coefficient can be well estimated as $\Delta^2\nu_B$. When the two resonant regions are separated more apart, the diffusion coefficient drops as the system becomes more regular. When $\omega_B = 4$, the diffusion coefficient is approximately $\sim \Delta^2\nu_L$.

\begin{figure}[h]
\includegraphics[width=6in, height=2.3in]{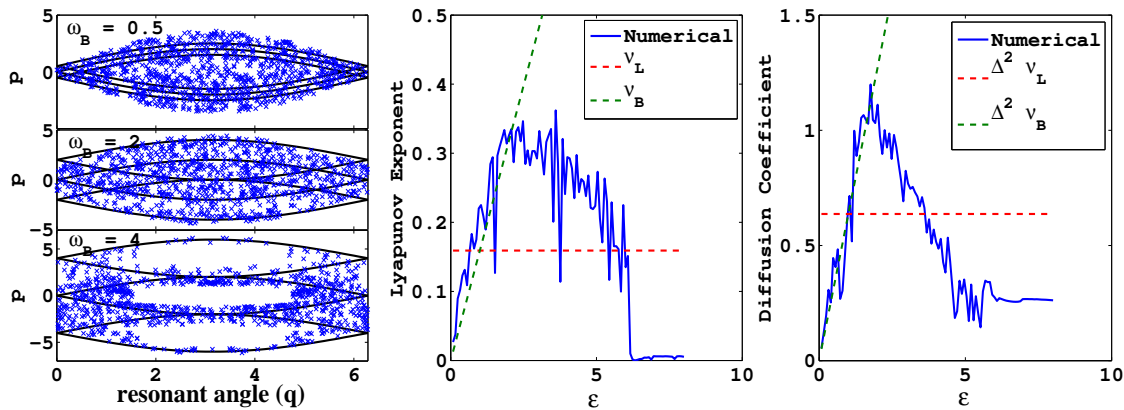}
\caption{\label{f:triple} The analytical models for the triple resonances analogous to the bridge zone. Left panel: the surface of section of the double resonances with different overlaps starting with $q=1$, $p=1.5$. Middle panel: the numerical and the analytical estimates of the Lyapunov exponent with different overlaps. Right panel: the numerical and analytical estimates of the diffusion coefficient with different overlaps.}
\vspace{0.1cm}
\end{figure}

Similarly, for the triple resonances, we use the following simplified Hamiltonian:
\begin{eqnarray}
H_T(p, q, t) = \frac{\beta}{2}p^2 + c(\cos q + \cos(q - \omega_B t) + \cos(q + \omega_B t))
\end{eqnarray}
Using trigonometric identities, this Hamiltonian can be rewritten as 
\begin{eqnarray}
H_T(p, q, t) = \frac{\beta}{2}p^2 + c(1+2\cos(\omega_B t)) \cos q
\end{eqnarray}
Thus, $H_T$ can be understood as a ``breathing" resonance whose width is changing with frequency $\nu_B$ \citep{Morbidelli02}. We plot the overlap of the resonances in the left panel of Figure (\ref{f:triple}). 

We numerically calculated the Lyapunov exponent and the diffusion coefficient with the method described for the double resonances. We find that similar to the double resonances, the Lyapunov exponent can be well estimated as $\nu_B$ when $\omega_B < 2$, as $\sim 2 \nu_L$ when $2<\omega_B<4$ and drops when $\omega_B>4$. For the diffusion coefficient, we find that it can be estimated as $\Delta^2 \nu_B$ for $\omega_B<2$ and it drops for $\omega_B>2$. At $\omega_B \sim 4$, it can be estimated as $\Delta^2 \nu_L$.






\end{document}